# Solar wind speed theory and the nonextensivity of solar corona

Du Jiulin[*], Song Yeli

*Department of Physics, School of Science, Tianjin University, Tianjin 300072, China*

**Abstract.** The solar corona is a complex system, with nonisothermal plasma and being in the self-gravitating field of the Sun. So the corona plasma is not only a nonequilibrium system but also a nonextensive one. We estimate the parameter of describing the degree of nonextensivity of the corona plasma and study the generalization of the solar wind speed theory in the framework of nonextensive statistical mechanics. It is found that, when use Chapman's corona model (1957) as the radial distribution of the temperature in the corona, the nonextensivity reduces the gas pressure outward and thus leads a significant deceleration effect on the radial speed of the solar wind.

---

[*] Email Address: jiulindu@yahoo.com.cn

# 1. Introduction

Solar corona is a complex system. The solar wind flow (coronal outflow) is due to the huge difference in the pressure between the solar corona and the interstellar space. This pressure difference drives plasma outward despite the restraining influence of the solar gravitation. Traditionally, the gas of corona plasma is assumed naturally to follow from the kinetic theory in the Maxwellian sense, where the pressure is taken to be the one of an ideal gas (Hundhausen, 1972). However, due to the long-range nature of the gravitating interactions, the gas under the field of solar gravitation behaves *nonextensively*, which would lead to the particles not strictly to follow Maxwellian distribution but, in some situations, to be power-law one (Cranmer, 1998; Lavagno and Quarati, 2006; Du, 2004a, 2006a, 2007). Let us now study the nonextensive effect of the corona plasma in the gravitational field of the Sun on the solar wind speed. We will consider this system in terms of the theory of astrophysical fluid dynamics and introduce the nonextensive effect into the solar wind theory.

In the conventional theory for the solar wind speed, the velocity distribution function for the particles has been usually assumed to be the Maxwellian one, which has been incarnated in the theory by the use of the equation of ideal gas as well as the Boltzmann-Gibbs entropy since the basic theory of solar wind was proposed by Parker in 1958. However, observations of space plasmas seldom indicate the presence of ideal Maxwellian velocity distributions. Spacecraft measurements of plasma velocity distributions in the solar wind have revealed that non-Maxwellian distributions are quite common (see, e.g., Cranmer 1998 and the references in). In many situations the distributions appear reasonably Maxwellian at low energies but have a "suprathermal" power-law tail at high energies. This has been well modeled by the $\kappa$-distribution (Olbert 1969; Collier et al.1996), one statistical distribution equivalent to the *q*-distribution presented now in the framework of nonextensive statistical mechanics (Leubner, 2002; Leubner and Voros, 2004; Du, 2004a, 2004b, 2006a, 2006b, 2006c; Lima et al, 2000; Silva et al, 1998) by the power-law form,

$$f(v) = B_q \left[ 1 - (1-q) \frac{mv^2}{2kT} \right]^{\frac{1}{1-q}}, \tag{1}$$

It has been found that the nonextensivity of the systems with slef-gravitating long-rang interactions, in certain situations, can be described by the above power-law $q$-distribution. In the solar corona, the $\kappa$-like distributions have been proposed to arise from strong nonequilinrium thermodynamic gradients, Fermi acceleration at upwelling convection-zone waves or shocks, and electron-ion runaway in a Dreicerorder electric field (see, e.g., Cranmer 1998 and the references in).

We want to know such a $\kappa$-like distribution will produce relatively to the Maxwellian one what effects on the speed of the solar wind rather than to explain the phenomena observed. For this purpose, we first in Sec.2 simply review the conventional theory for the solar wind speed, and then in Sec.3 we study the degree of nonextensivity of the solar corona plasma. In Sec.4, we generalize the theory in the framework of nonextensive statistics and in Sec.5 we investigate the nonextensive effect on the solar wind speed. Finally in Sec.6, we give the conclusive remarks.

**2. The conventional theory for the solar wind speed**

As we know that the solar wind plasma is believed to be a nonequlibrium system and the atmosphere of the Sun is assumed to be steady and spherically symmetric. In such situations, the basic fluid dynamical equations in the solar wind theory are given ( Zhang, 1992; Hundhausen, 1972) by the mass equation (the continuity equation),

$$4\pi r^2 \rho v = \text{constant}, \qquad (2)$$

the momentum equation,

$$\rho v \frac{dv}{dr} = -\frac{dP}{dr} - \frac{GM}{r^2}\rho + \rho D, \qquad (3)$$

and the energy equation,

$$\rho v \frac{ds}{dr} = \frac{Q}{T}, \qquad (4)$$

where $r$ is the distance from the center of the Sun; $\rho, T, P$ and $s$ are the density, the temperature, the pressure and the specific entropy, respectively; $v$ is the speed of the fluid, $Q$ is the rate of the internal energy generated by the nonradiative processes (including the heat transfer), and $D$ is the rate of the momentum obtained by the ways except the heat pressure gradient and the gravity. Usually, the pressure of the atmosphere of the Sun takes the form of the equation of state of ideal gas, $P = \rho kT/m$ based on the standard statistical mechanics, with $m$ the mass of the proton and $k$ the Boltzmann constant and the specific entropy takes the form,

$$s = \frac{3}{2}\frac{k}{m}\ln(P\rho^{-5/2}) + \text{const.} \tag{5}$$

In the literatures that discuss the solar wind theory, one usually makes use of Eq.(5) and combine Eqs.(3) and (4) to derive the differential form of the equations,

$$\frac{d}{dr}\left(\frac{1}{2}v^2 + \frac{5}{2}\frac{kT}{m} - \frac{GM}{r}\right) = \frac{Q}{\rho v} + D. \tag{6}$$

This form can be used to explain why the energy flux brought by the steady convection of the plasma may be changed by the local heating to produce the heat pressure gradient or by directly increasing the momentum, such as the wave pressure gradient. If delete $\rho$ from Eqs.(2) and (3), we can obtain the so-called continuity momentum equation,

$$\left(v - \frac{v_s^2}{v}\right)\frac{dv}{dr} = \frac{2v_s^2}{r} - \frac{dv_s^2}{dr} - \frac{GM}{r^2} + D, \tag{7}$$

where $v_s = \sqrt{kT/m}$ is the sound speed. Eq.(7) tells us that the corona plasma might be accelerated by the combination effects of those terms on the right hand side of Eq.(7). Here we will introduce the nonextensivity into the above theory. It might be one possible acceleration effect on the solar wind speed.

**3. Temperature, pressure and nonextensivity of the corona**

The solar corona is a plasma system far from equilibrium. According to Chapman's corona model (1957), the corona temperature varies dependent on the radial distance from the center of the Sun (Lin, 2000; Zhang, 1992) as follows,

$$T(r) = T_0\left(\frac{r_0}{r}\right)^{2/7}, \quad r \geq r_0, \tag{8}$$

where $r_0$ is a reference distance from the center of the Sun. For example, it may be taken as $r_0 = 1.05R_\odot$, $R_\odot$ is the radius of the Sun. $T_0$ is the temperature at $r = r_0$. The dependence of pressure on the radial distance is expressed by

$$P(r) = P_0 \exp\left(-\frac{GM\mu}{\Re} \int_{r_0}^{r} \frac{dr'}{r'^2\, T(r')}\right), \tag{9}$$

where $P_0$ is the pressure at $r = r_0$; $G, M, \Re$ and $\mu$ are the gravitational constant, the mass of the Sun, the gas constant and the mean molecular weight, respectively.

We now consider the nonextensivity of the solar corona. The nonextensivity is one non-additive property of the nonequlibrium system being an external field with the long-rang inter-particle interactions. We have already known examples of this in the many-body treatment of self-gravitating systems and plasma systems. The degree of the nonextensivity for a nonequilibrium system with the long-range inter-particle interactions such as the gravitational force can be estimated by using the nonextensive parameter $q$ defined in nonextensive statistical mechanics (NSM) in terms of its deviation from unity. The formulation of the parameter $q$ for the self-gravitating system can be given (Du, 2004a; 2006a) by

$$1 - q = -\frac{k}{\mu m_H} \frac{dT}{dr} \bigg/ \frac{GM}{r^2}, \tag{10}$$

where $k$ is the Boltzmann constant and $m_H$ is the mass of hydrogen atom. The deviation of $q$ from unity is thought to describe the degree of nonextensivity. If substituting Eqs.(8) into Eq.(10), we have clearly understood that *the solar corona is not only a nonequilibrium system but also a nonextensive system*. Therefore, we have to generalize the solar wind speed theory in the framework of NSM so as to take the nonextensive effect into consideration, though the conventional theory has made very extensive applications.

From Eq.(8) we can find the temperature gradient in the corona,

$$\frac{dT}{dr} = -\frac{2}{7} T_0 \left(\frac{r_0}{r}\right)^{2/7} \frac{1}{r}, \tag{11}$$

and then, make use of (10),

$$1 - q = \frac{2k T_0}{7m} \left(\frac{r_0}{r}\right)^{2/7} \bigg/ \frac{GM}{r}, \tag{12}$$

where $m$ is the mass of proton. When we consider the case of $r_0 = 1.05 R_\odot$ and $T_0=1.8\times10^6$K (Zhang, 1992), the temperature gradients and the values of 1-$q$ at the different distances from the sun can be obtained from Eqs.(11) and (12). They are shown in Table I. We find that the value of 1-$q$ (the degree of nonextensivity) in the corona rises with the increase of the radial distance $r$ from the Sun, though the temperature gradient slopes more and more gently. The farther the radial distance is from the Sun, the higher the degree of nonextensivity is, which, as we will see, with the distances far away from the Sun, would have a more significant deceleration effect on the solar wind's speed.

Table I. The values of 1-$q$ in the solar corona

| $r/r_0$ | $T/T_0$ | $dT/dr$ (K /km) | 1-$q$ |
|---|---|---|---|
| 1   | 1.00  | -0.7037 | 0.0234 |
| 2   | 0.82  | -0.2886 | 0.0384 |
| 4   | 0.67  | -0.1184 | 0.0631 |
| 6   | 0.60  | -0.0703 | 0.0843 |
| 8   | 0.55  | -0.0486 | 0.1036 |
| 10  | 0.52  | -0.0364 | 0.1200 |
| 15  | 0.46  | -0.0216 | 0.1618 |
| 20  | 0.42  | -0.0149 | 0.1984 |
| 30  | 0.38  | -0.0089 | 0.2667 |
| 40  | 0.35  | -0.0061 | 0.3267 |
| 50  | 0.33  | -0.0046 | 0.3832 |
| 60  | 0.31  | -0.0036 | 0.4365 |
| 70  | 0.297 | -0.0030 | 0.4873 |
| 80  | 0.286 | -0.0025 | 0.5360 |
| 90  | 0.276 | -0.0022 | 0.5831 |
| 100 | 0.268 | -0.0019 | 0.6294 |

**4. A generalization of the solar wind speed theory in NSM**

The nonextensive statistical mechanics based on Tsallis entropy can be defined by the so-called $q$-logarithmic function, $\ln_q f$, and $q$-exponential function, $\exp_q f$, (for example, Tsallis et al, 1998; Lima et al, 2001)

$$\ln_q f = (1-q)^{-1}(f^{1-q}-1), \quad f>0; \tag{13}$$

$$\exp_q f = [1+(1-q)f]^{1/1-q}, \tag{14}$$

with 1+(1-q)f >0 and $\exp_q f = 0$ otherwise, where $q$ is the nonextensive parameter different from unity. In other words, the deviation of $q$ from unity describes the nonextensive degree of the system under consideration. When $q \to 1$ all the above expressions reproduce those verified by the usual elementary functions, and Tsallis entropy function, $S_q = -k \int f^q \ln_q f \, d\Omega$, reduces to the standard Boltzmann-Gibbs logarithmic one, $S = -k \int f \ln f \, d\Omega$. In this new statistical framework, the pressure of the atmosphere of the Sun can be written (Du, 2006a, 2006c) as

$$P_q = C_q \rho kT / m \tag{15}$$

with the $q$-coefficient $C_q = 2/(5-3q)$, $0<q<5/3$. And naturally, the specific entropy can be written by

$$s_q = \frac{3}{2}\frac{k}{m} \ln_q (P_q \rho^{-5/2}) + \text{constant}, \tag{16}$$

After the $q$-logarithmic function is replaced by Eq.(13), one has

$$s_q = \frac{3}{2}\frac{k}{m}\frac{1}{1-q}\left[P_q^{1-q}\rho^{-5(1-q)/2} - 1\right] + \text{constant}. \tag{17}$$

We use Eqs.(17) and (15) instead of Eq.(5) and the equation of state of ideal gas, respectively, then in the new framework, Eq.(6) and Eq.(7) become

$$\frac{d}{dr}\left(\frac{1}{2}v^2 + \frac{5C_q}{2}\frac{kT}{m} - \frac{GM}{r}\right) = C_q^{2-q}\left(\frac{kT}{m}\right)^{1-q}\rho^{-(5-3q)/2}\frac{Q}{v} + D, \tag{18}$$

and

$$\left(v - C_q\frac{v_s^2}{v}\right)\frac{dv}{dr} = C_q\frac{2v_s^2}{r} - C_q\frac{dv_s^2}{dr} - \frac{GM}{r^2} + D. \tag{19}$$

Thus the nonextensivity of the solar corona plasma has been introduced into the theory of the solar wind. These new equations tell us that the nonextensive parameter $q$ different from unity will play a role in the acceleration of the solar wind speed.

**5. The nonextensive effect on the solar wind speed**

In Parker's theory of solar wind speed (1958), the solar corona is assumed approximately to be isothermal one and the nonextensive effect is neglected because the temperature gradient is thought to be zero. Thus, the fluid dynamical equations are the equation of the conservation of mass, Eq.(2), and the equation of the conservation of momentum, Eq.(3), while the energy equation, Eq.(4), is discarded. In such a case, the dynamical equations for the solar wind speed become directly Eq.(7). Namely,

$$\left(v - \frac{v_s^2}{v}\right)\frac{dv}{dr} = \frac{2v_s^2}{r} - \frac{GM}{r^2}. \tag{20}$$

We now consider the nonextensive effects expressed by Eq.(10) on the Parker's speed theory. In the present case that the nonextensive effect is taken into consideration, the dynamical equation for the solar wind speed should be replaced by the generalized one, Eq.(19). Namely,

$$\left(v - C_q \frac{v_s^2}{v}\right)\frac{dv}{dr} = C_q \frac{2v_s^2}{r} - C_q \frac{dv_s^2}{dr} - \frac{GM}{r^2}. \tag{21}$$

If $q \to 1$ it reduces to the conventional form, Eq.(20). In the new dynamical equation, the new critical speed for the solar wind is no longer the sound speed but is $v_c(q) = v_s\sqrt{C_q} = \sqrt{2C_q kT/m}$. The new critical distance $r_c(q)$ can be determined by the equation

$$r_c^2(q) - \frac{2T}{dT/dr}r_c(q) + \frac{2T}{dT/dr}\frac{r_c(1)}{C_q} = 0, \tag{22}$$

where $r_c(1) = r_c(q=1) = GM/2v_s^2$ is the old critical distance in the Parker' theory when the nonextensive effect is not considered. When $r$ is very large, we consider $dT/dr \to 0$, then we get

$$r_c(q) = \frac{GMm}{4C_q kT} = \frac{r_c(1)}{C_q}. \tag{23}$$

and Eq.(21) becomes

$$\left[\left(\frac{v}{v_c(q)}\right)^2 - 1\right]\frac{dv}{v} = 2\left[1 - \frac{r_c(1)}{C_q r}\right]\frac{dr}{r} - 2d\ln v_s, \tag{24}$$

Complete the integral of above equation, we find

$$\left(\frac{v}{v_c(q)}\right)^2 - 2\ln v = 4\ln r + \frac{4r_c(1)}{C_q r} - 4\ln v_s + const. \tag{25}$$

where the integral constant can be determined by using $r = r_c(q)$, $v = v_c(q)$. Thus

Eq.(25) becomes

$$\left(\frac{v}{v_c(q)}\right)^2 - \ln\left(\frac{v}{v_c(q)}\right)^2 = 4\ln\frac{r}{r_c(q)} + \frac{4r_c(1)}{C_q}\left(\frac{1}{r} - \frac{1}{r_c(q)}\right) + 1, \quad (26)$$

It determines variations of the radial speed dependent on the radial distance from the Sun. It is clear that all the equations depend explicitly on the nonextensive parameter $q$. When $q \to 1$, Eq.(26) recovers perfectly the speed equation of Parker's solar wind theory under the assumption of isothermal corona (Zhang, 1992),

$$\left(\frac{v}{v_c(1)}\right)^2 - \ln\left(\frac{v}{v_c(1)}\right)^2 = 4\ln\frac{r}{r_c(1)} + \frac{4r_c(1)}{r} - 3, \quad (27)$$

where $v_c(1) = v_s = \sqrt{2kT/m}$.

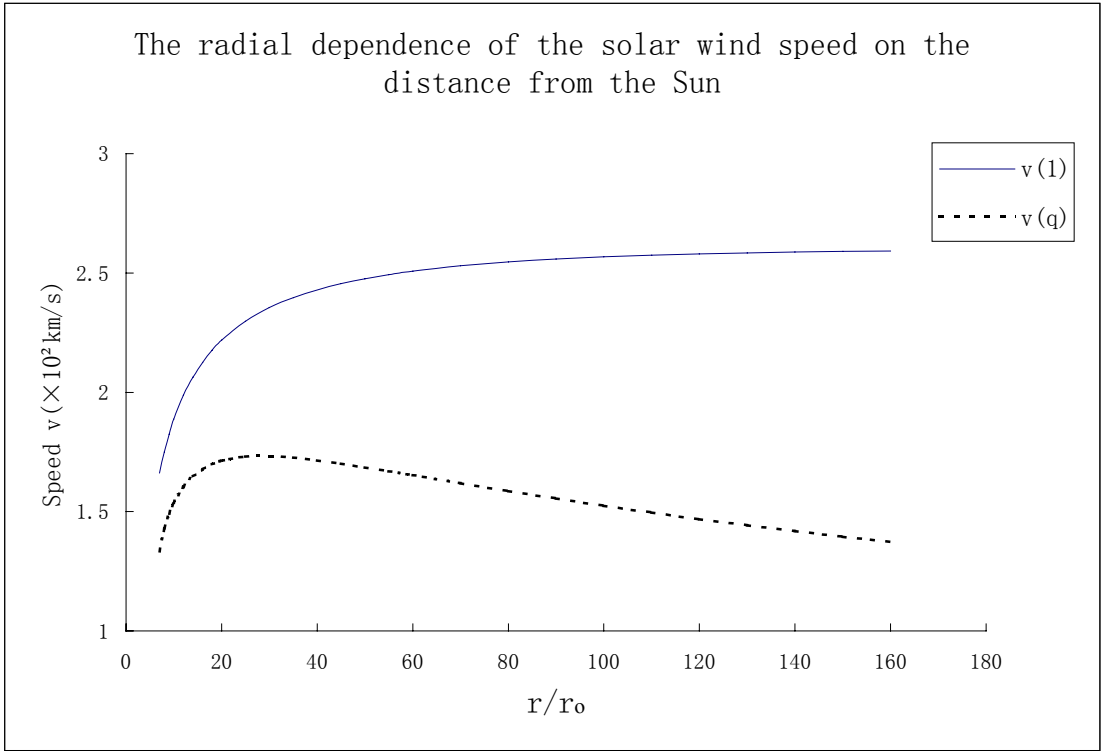

Fig.I. The radial variation of the solar wind speed dependent on the distance from the center of the Sun. The dashed line denotes the speed in the case that the nonextensive effects are taken into account. The solid line denotes the Parker's speed but considering the radial variation of the temperature.

By the numerical calculations, the radial dependence of the solutions of Eq.(26) and Eq.(27) are obtained, respectively. They are shown in Fig.I The dashed line denotes the solution of Eq.(26), being the solar wind speed in the case that the nonextensive effects are taken into account. The solid line denotes the solution of Eq.(27), being the Parker's solar wind speed but considering the radial variation of the temperature. These results clearly show that the nonextensivity of the solar corona has a significant deceleration effect on the radial speed of solar wind. And this effect would be enhanced with the distance far away from the Sun.

From Eq.(26) we can derive the speed formula of the solar wind at the distances far away from the Sun. For instance, when $r$ is very large, we have $4r_c(1)/C_q r \to 0$ and, by using Eq.(23), $-4r_c(1)/C_q r_c(q) = -4$. Then $-\ln(v/v_c(q))^2$ and -3 are very small as compared with the terms, $(v/v_c(q))^2$ and $4\ln(r/r_c(q))$, and so may be neglected. Thus Eq.(26) can be reduced to

$$v^2(q) \approx 4v_c^2(q)\left[\ln\frac{r}{r_c(q)}\right] . \tag{28}$$

It is clear that when $q \to 1$ the speed formula of Parker's solar wind theory for $r$ to be very large, $v(1) \approx 2v_s\left[\ln r/r_c(1)\right]^{1/2}$ (Zhang, 1992; Hundhausen, 1972), can also be recovered correctly from Eq.(28). Substitute the related parameters into Eq.(28), we find

$$v^2(q) = 4C_q v_s^2\left[\ln\frac{r}{r_c(q)}\right]$$
$$= \frac{2}{5-3q}\left[v^2(1) + \frac{8kT}{m}\ln\frac{2}{5-3q}\right]. \tag{29}$$

By using this formula we can calculate the square of speed of the solar wind far away from the Sun in the case that the nonextensive effect is taken into consideration. For example, the radial speed of solar wind at $r = 100 r_0$ can be written as

$$v(q=0.37) = 0.72 v(1)\left(1 - \frac{4.12 \times 10^{14}}{v^2(1)}\right)^{1/2} \text{ cm s}^{-1}. \tag{30}$$

with the nonextensive parameter $q$=0.37. If take Parker's speed as $v(1)$=263 km s$^{-1}$, for example, then we find the new speed to become $v(q=0.37)$=157 km s$^{-1}$. This modification introduced by the nonextensive effects is quite significant.

## 6. Conclusive remarks

We have studied the degree of nonextensivity of the solar corona and have introduced the nonextensive effect into the theory for the solar wind speed. The nonextensivity is shown to be a possible deceleration effect on the radial speed of the solar wind. Our results are concluded as follows.

(a) The solar corona is not only a nonequilibrium system but also a nonextensive system. The degree of nonextensivity in the system rises with the increase of the radial distance from the Sun, though the temperature gradient slopes more and more gently.

(b) The nonextensivity of the solar corona has a significant deceleration effect on the radial speed of solar wind. And this effect will be enhanced with the increase of the distance away from the Sun.

(c) In the new framework of NSM, unlike the case in Parker's theory, the solar wind speed does not rise monotonously with the increase of the distance from the Sun, but rises rapidly with the distance in the regions not very far from the Sun ($< 20R_\odot$), being the maximum at about $r = 28r_0$. Further out from the Sun, beyond the maximum, the speed decreases very slowly with the distance away from the Sun.

(d) In Parker's speed equation Eq.(26), If we consider the dependence of the corona temperature $T$ on the distance $r$, as expressed by Eq.(7), all the calculated solar wind speeds at the different places are significantly less than those in Parker's solar wind speed theory, namely under the assumption of isothermal corona.

Why does the nonextensivity have the deceleration effect on the solar wind speed? This question may be answered from the role of the nonextensive effect in the gas pressure of the solar corona. We know that the nonextensive parameter $q$ is determined by Eq.(12) and its values are strongly depended on the radial distribution of the temperature in the solar corona. When we take Eq.(8) as the temperature model of the corona, the parameter $q$ is always less than unity. The introduction of such a nonextensive effect leads to a decrease of the gas pressure outward (see Eq.(15))and thus produces the deceleration of the solar wind speed.

**Acknowledgements**

This work is supported by the National Natural Science Foundation of China under the grant No.10675088 and the "985" Program of TJU of China. We deeply acknowledge H. J.Haubold for continuous encouragement and for his interest in our works on astrophysical applications of statistical mechanics. We thank S. Abe, A. M. Mathai and C. Tsallis for useful discussions.